\documentclass[10pt,journal]{IEEEtran}
\usepackage{tabularx,booktabs}
\usepackage{xcolor}
\usepackage{hyperref}

\ifCLASSOPTIONcompsoc
  \usepackage[nocompress]{cite}
\else
  \usepackage{cite}
\fi
\ifCLASSINFOpdf
\else
\fi
\usepackage{algorithm, amsmath}
\usepackage[noend]{algpseudocode}

\algnewcommand\algorithmicforeach{\textbf{for each}}
\algdef{S}[FOR]{ForEach}[1]{\algorithmicforeach\ #1\ \algorithmicdo}

\usepackage{subcaption}
\usepackage{graphicx}
\usepackage{url}

\newif\ifcolormarker
\colormarkertrue

\begin{document}
%
\title{Deep Reinforcement Learning meets Graph Neural Networks: exploring a routing optimization use case}
%
%
%
%

\author{Paul~Almasan, José~Suárez-Varela, Krzysztof~Rusek, Pere~Barlet-Ros, Albert~Cabellos-Aparicio
\IEEEcompsocitemizethanks{
\IEEEcompsocthanksitem Paul~Almasan, José~Suárez-Varela, Pere~Barlet-Ros and Albert~Cabellos-Aparicio are with the Barcelona Neural Networking Center. Universitat Politècnica de Catalunya. Barcelona, Spain.\protect\\
E-mail: \{felician.paul.almasan, jose.suarez-varela, pere.barlet, alberto.cabellos\}@upc.edu,
\IEEEcompsocthanksitem Krzysztof~Rusek is with the Institute of Telecommunications, AGH University of Science and Technology, Krakow, Poland, and with the Barcelona Neural Networking Center, Universitat Politècnica de
Catalunya, Barcelona, Spain. E-mail: krusek@agh.edu.pl \protect\\

\textbf{NOTE:} This work has been accepted for publication in the Computer Communications journal. Please use the following reference to cite this work: Paul Almasan, José Suárez-Varela, Krzysztof Rusek, Pere Barlet-Ros and Albert Cabellos-Aparicio. \textit{"Deep reinforcement learning meets graph neural networks: Exploring a routing optimization use case"} in Computer Communications, 2022, doi: \href{https://doi.org/10.1016/j.comcom.2022.09.029}{https://doi.org/10.1016/j.comcom.2022.09.029}.
}
}

\IEEEtitleabstractindextext{%
\begin{abstract}
Deep Reinforcement Learning (DRL) has shown a dramatic improvement in decision-making and automated control problems. Consequently, DRL represents a promising technique to efficiently solve many relevant optimization problems (e.g., routing) in self-driving networks. However, existing DRL-based solutions applied to networking fail to generalize, which means that they are not able to operate properly when applied to network topologies not observed during training. This lack of generalization capability significantly hinders the deployment of DRL technologies in production networks. This is because state-of-the-art DRL-based networking solutions use standard neural networks (e.g., fully connected, convolutional), which are not suited to learn from information structured as graphs. 

In this paper, we integrate Graph Neural Networks (GNN) into DRL agents and we design a problem specific action space to enable generalization. GNNs are Deep Learning models inherently designed to generalize over graphs of different sizes and structures. This allows the proposed GNN-based DRL agent to learn and generalize over arbitrary network topologies. We test our DRL+GNN agent in a routing optimization use case in optical networks and evaluate it on 180 and 232 unseen synthetic and real-world network topologies respectively. The results show that the DRL+GNN agent is able to outperform state-of-the-art solutions in topologies never seen during training.
\end{abstract}

\vspace{0.9cm}
}

\maketitle

\IEEEdisplaynontitleabstractindextext

%
\IEEEpeerreviewmaketitle

\IEEEraisesectionheading{\section{Introduction}\label{sec:introduction}}

%
%
%
%
In the last years, industrial advances (e.g., Industry 4.0, IoT) and changes in social behavior created a proliferation of modern network applications (e.g., Vehicular Networks, AR/VR, Real-Time Communications), imposing new requirements on backbone networks (e.g., high throughput and low latency). Consequently, network operators need to efficiently manage the network resources, ensuring the customer’s Quality of Service and fulfilling the Service Level Agreements. This is typically done using expert knowledge or solvers leveraging Integer Linear Programming (ILP) or Constraint Programming (CP). However, real-world production networks have in the order of hundreds of nodes and solvers based on ILP or CP  would take a large amount of time to solve network optimization problems \cite{hartert2015declarative, knight2011internet}. In addition, heuristic based solutions are far from being optimal.

Deep Reinforcement Learning (DRL) has shown significant improvements in sequential decision-making and automated control problems~\cite{mnih2015human, silver2017mastering}. As a result, the network community is already investigating DRL as a key technology for network optimization (e.g., routing) with the goal of enabling \mbox{self-driving} networks~\cite{feamster2017and, wang2017machine,mestres2017knowledge,kalmbach2018empowering}. However, existing DRL-based solutions still fail to~\emph{generalize} when applied to different network scenarios \cite{valadarsky2017learning,rusek2020routenet}.
In this context, generalization refers to the ability of the DRL agent to adapt to new network scenarios not seen during training (e.g., network topologies, configurations).

We argue that generalization is an essential property for the successful adoption of DRL technologies in production networks.  Without generalization, DRL solutions should be trained in the same network where they are deployed, which is not possible or affordable in general. To train a DRL agent is a very costly and lengthy process. It often requires significant computing power and instrumentation of the network to observe its performance (e.g., delay, jitter). Additionally, decisions made by a DRL agent during training can lead to degraded performance or even to service disruption. Thus, training a DRL agent in the customer's network may be unfeasible.

With generalization, a DRL agent can be trained with multiple, representative network topologies and configurations. Afterwards, it can be applied to other topologies and configurations, as long as they share some common properties. Such a ``universal'' model can be trained in a laboratory and later on be incorporated in a product or a network device (e.g., router, load balancer). The resulting solution would be ready to be deployed to a production network without requiring any further training or instrumentation in the customer network\footnote{Note that solutions based on transfer learning do not offer this property as DRL agents need to be re-trained on the network where they finally operate.}. 

Unfortunately, existing DRL proposals for networking were designed to operate in the same network topology seen during training~\cite{chen2018deep, valadarsky2017learning, xu2018experience}, thereby limiting their potential deployment on production networks. 
The main reason behind this strong limitation is that computer networks are fundamentally represented as graphs. For instance, the network topology and routing policy are typically represented as such. However, state-of-the-art proposals~\cite{chen2018deep, chen2018auto, mestres2018understanding, suarez2019routingJ} use traditional neural network (NN) architectures (e.g., fully connected, convolutional) that are not well suited to model graph-structured information~\cite{battaglia2018relational}.

In this paper, we integrate Graph Neural Networks (GNN)~\cite{scarselli2008graph} into DRL agents to solve network optimization problems. Particularly, our architecture is intended to solve routing optimization in optical networks and to generalize over never-seen arbitrary topologies. The GNN integrated in our DRL agent is inspired by Message-passing Neural Networks (MPNN), which were successfully applied to solve a relevant chemistry-related problem~\cite{gilmer2017neural}. In our case, the GNN was specifically designed to capture meaningful information about the relations between the links and the traffic flowing through the network topologies.

The evaluation results show that our agent achieves a strong generalization capability compared to state-of-the-art DRL (SoA DRL) algorithms\cite{suarez2019routingJ}. Additionally, to further test the generalization capability of the proposed DRL-based architecture, we evaluated it in a set with 232 different real-world network topologies. The results show that the proposed DRL+GNN architecture is able to achieve outstanding performance over the networks never seen during training. Finally, we explore the generalization limitations of our architecture and discuss its scalability properties. 

Overall, our DRL+GNN architecture for network optimization has the following features:

\begin{itemize}
    \item {\em Generality:} It can work effectively in network topologies and scenarios never seen during training.
    \item {\em Deployability:} It can be deployed to production networks without requiring training nor instrumentation in the customer network.
    \item {\em Low overhead:} Once trained, the DRL agent can make routing decisions in only one step ($\approx$ \textit{ms}), while its cost scales linearly with the network size.
    \item {\em Commercialization:} Network vendors can easily embed it in network devices or products, and successfully operate "arbitrary" networks.
\end{itemize}

We believe the combination of these features can enable the development of a new generation of networking solutions based on DRL that are more cost-effective than current approaches based on heuristics or linear optimization. All the topologies and scripts used in the experiments, as well as the source code of our DRL+GNN agent are publicly available~\cite{code}.

\section{Background}
\label{section:back}

The solution proposed in this paper combines two machine learning mechanisms. First, we use a GNN to model computer network scenarios. GNNs 
are neural network architectures specifically designed to generalize over graph-structured data~\cite{battaglia2018relational}. In addition, they offer near real-time operation in the scale of milliseconds (see Section \ref{subsec:computime}). Second, we use Deep Reinforcement Learning to build an agent that learns how to efficiently operate networks following a particular optimization goal. DRL applies the knowledge obtained in past optimizations to later decisions, without the necessity to run computationally intensive algorithms.

\subsection{Graph Neural Networks}

Graph Neural Networks are a novel family of neural networks designed to operate over graphs. They were introduced in \cite{scarselli2008graph} and numerous variants have been developed since \cite{li2015gated, velivckovic2017graph}. In their basic form, they consist of associating some initial states to the different elements of an input graph, and combining them considering how these elements are connected in the graph. An iterative algorithm updates the elements' state and uses the resulting states to produce an output. The particularities of the problem to solve will determine which GNN variant is more suitable, depending on, for instance,  the nature of the graph elements (i.e., nodes and edges) involved.

Message Passing Neural Networks (MPNN) \cite{gilmer2017neural} are a well-known type of GNNs that apply an iterative message-passing algorithm to propagate information between the nodes of the graph. In a message-passing step, each node \textit{k} receives messages from all the nodes in its neighborhood, denoted by \textit{N(k)}. Messages are generated by applying a message function \textit{m(·)} to the hidden states of node pairs in the graph. Then, they are combined by an aggregation function, for instance, a sum (Equation \ref{eq:message_function}). Finally, an update function \textit{u(·)} is used to compute a new hidden state for every node (Equation \ref{eq:update_function}).

\begin{equation}
M_{k}^{t+1} = \sum_{i \in N(k)} m(h_{k}^{t}, h_{i}^{t})
\label{eq:message_function}
\end{equation}

\begin{equation}
h_{k}^{t+1} = u(h_{k}^{t}, M_{k}^{t+1})
\label{eq:update_function}
\end{equation}

Where functions \textit{m(·)} and \textit{u(·)} can be learned by neural networks. After a certain number of iterations, the final node states are used by a readout function \textit{r(·)} to produce an output for the given task. This function can also be implemented by a neural network and is typically tasked to predict properties of individual nodes (e.g., the node's class) or global properties of the graph.

GNNs have been able to achieve relevant performance results in multiple domains where data is typically structured as a graph \cite{gilmer2017neural, battaglia2016interaction}. Since computer networks are fundamentally represented as graphs, it is inherent in their design that GNNs offer unique advantages for network modeling compared to traditional neural network architectures (e.g., fully connected NN, Convolutional NN, etc.). 

\subsection{Deep Reinforcement Learning}

DRL algorithms aim at learning a long-term strategy that leads to maximize 
an objective function in an optimization problem. DRL agents start from a \textit{tabula rasa} state and they learn the optimal strategy by an iterative process that explores the state and action spaces. These are denoted by a set of states ($\mathcal{S}$) and a set of actions ($\mathcal{A}$). Given a state \textit{s $\in$ $\mathcal{S}$}, the agent will perform an action \textit{a $\in$ $\mathcal{A}$} that produces a transition to a new state \textit{s' $\in$ $\mathcal{S}$}, and will provide the agent with a reward \textit{r}. Then, the objective is to find a strategy that maximizes the cumulative reward by the end of an episode. The definition of the end of an episode depends on the optimization problem to address. 

Q-learning \cite{watkins1992q} is a RL algorithm whose goal is to make an agent learn a policy $\pi$ : $\mathcal{S}$ $\,\to\,$ $\mathcal{A}$. The algorithm creates a table (a.k.a., q-table) with all the possible combinations of states and actions. At the beginning of the training, the table is initialized (e.g., with zeros or random values) and during training, the agent updates these values according to the rewards obtained after selecting an action. These values, called q-values, represent the expected cumulative reward after applying action \textit{a} from state \textit{s}, assuming that the agent follows the current policy $\pi$ during the rest of the episode. During training, q-values are updated using the Bellman equation (see Equation \ref{eq:bellman}) where \textit{Q($s_t$,$a_t$)} is the q-value function at time-step \textit{t}, $\alpha$ is the learning rate, \textit{r($s_t$,$a_t$)} is the reward obtained from selecting action \textit{$a_t$} from state \textit{$s_t$} and $\gamma\in[0,1]$ is the discount factor.

\begin{equation}
\begin{split}
Q(s_t,a_t) = Q (s_t, a_t)+ \alpha\biggl( r(s_t,a_t) + \\ \gamma \max_{a'} Q (s_t',a') - Q(s_t, a_t) \biggr)
\end{split}
\label{eq:bellman}
\end{equation}

Deep Q-Network (DQN)~\cite{mnih2013playing} is a more advanced algorithm based on Q-learning that uses a Deep Neural Network (DNN) to approximate the q-value function. As the q-table size becomes larger, Q-learning faces difficulties to learn a policy from high dimensional state and action spaces. To overcome this problem, they proposed to use a DNN as a q-value function estimator, relying on the generalization capabilities of DNNs to estimate the q-values of states and actions unseen in advance. For this reason, a NN well suited to understand and generalize over the input data of the DRL agent is crucial for the agents to perform well when facing states (or environments) never seen before. Additionally, DQN uses an experience replay buffer to store past sequential experiences (i.e., stores tuples of \textit{\{s,a,r,s'\}} ).

\section{Network optimization scenario}
\label{section:optimi}

In this paper, we explore the potential of a GNN-based DRL agent to address the routing problem in Optical Transport Networks (OTN). Particularly, we consider a network scenario based on Software-Defined Networking, where the DRL agent (located in the control plane) has a global view of the current network state, and has to make routing decisions on every traffic demand as it arrives. We consider a traffic demand as the volume of traffic sent from a source to a destination node. This is a relevant optimization scenario that has been studied in the last decades in the optical networking community, where many solutions have been proposed~\cite{chen2018deep,suarez2019routingJ, kuri2003diverse}.

In our OTN scenario, the DRL agent makes routing decisions at the electrical domain, over a logical topology where nodes represent Reconfigurable Optical Add-Drop Multiplexers (ROADM) and edges are predefined lightpaths connecting them (see Figure~\ref{fig:drlenv}). The DRL agent receives traffic demands with different bandwidth requirements defined by the tuple $\{src, dst, bandwidth\}$, and it has to select an end-to-end path for every demand. Particularly, end-to-end paths are defined as sequences of lightpaths connecting the source and destination of a demand. Since the agent operates at the electrical domain, traffic demands are defined as requests of Optical Data Units (ODUk), whose bandwidth requirements are defined in the ITU-T Recommendation G.709~\cite{itu}. The ODUk signals are then multiplexed into Optical Transport Units (OTUk), which are data frames including Forward Error Correction. Eventually, OTUk frames are mapped to different optical channels within the lightpaths of the topology. 

In this scenario, the routing problem is defined as finding the optimal routing policy for each incoming source-destination traffic demand. The learning process is guided by an objective function that aims to maximize the traffic volume allocated in the network in the long-term. We consider that a demand is properly allocated if there is enough available capacity in all the lightpaths forming the end-to-end path selected. Note that lightpaths are the edges in the logical topology where the agent operates. The demands do not expire, occupying the lightpaths until the end of a DRL episode. This implies a challenging task for the agent, since it has not only to identify critical resources on networks (e.g., potential bottlenecks), but also to deal with the uncertainty in the generation of future traffic demands. The following constraints summarize the traffic demand routing problem in the OTN scenario:

\begin{figure}[!t]
  \centering
  \includegraphics[width=0.9\linewidth]{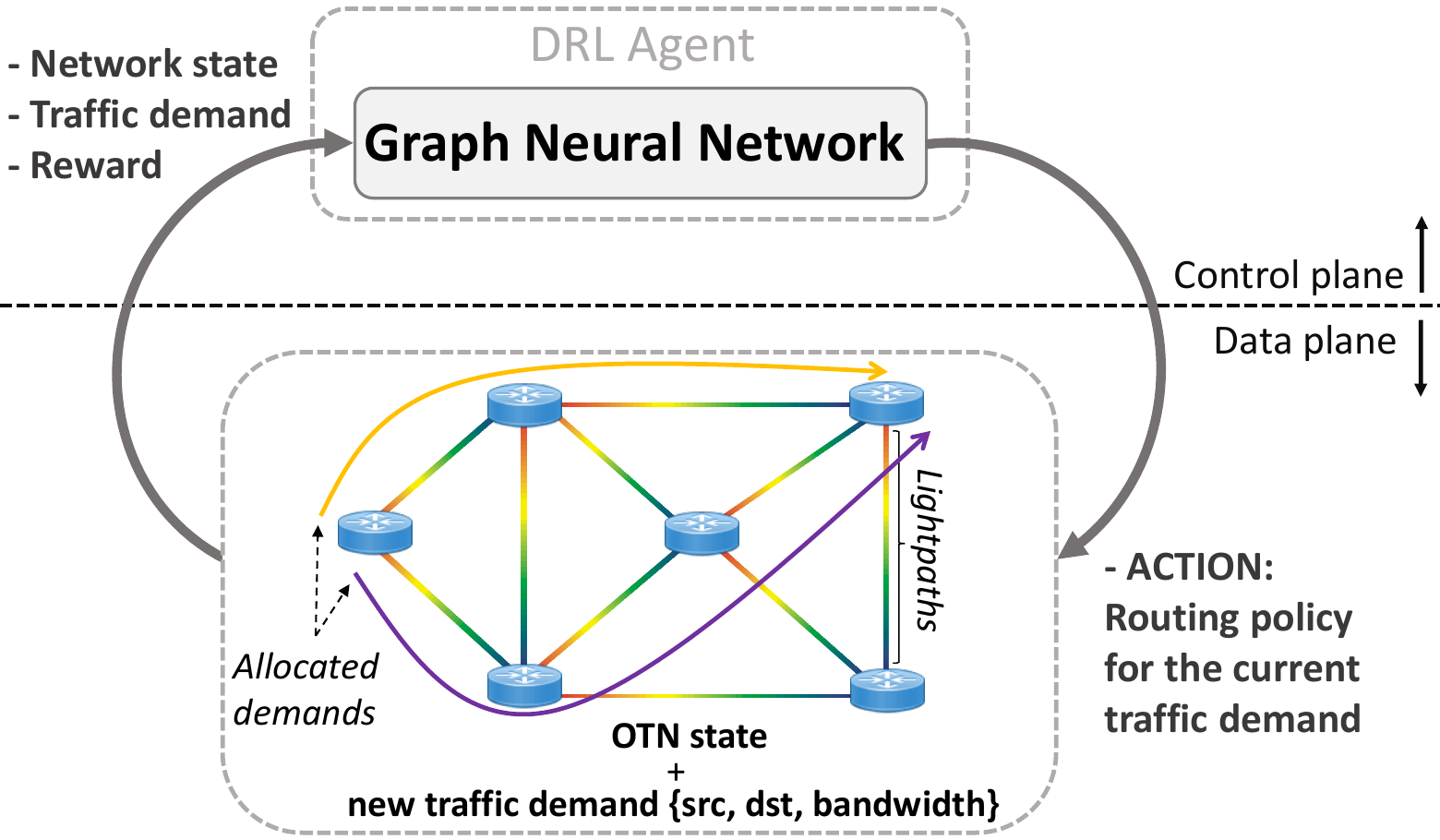}
  \caption{Schematic representation of the DRL agent in the OTN routing scenario.}
  \label{fig:drlenv}
\end{figure}

\begin{itemize}
    \item The agent must make sequential routing decisions for every incoming traffic demand
    \item Traffic demands can not be split over multiple paths
    \item Previous traffic demands can not be rerouted and they occupy the links' capacities until the end of the episode
\end{itemize}

The \emph{optimal} solution to the OTN optimization problem can be found by solving its Markov Decision Process (MDP)\cite{sutton2018reinforcement}. To do this, we can use techniques such as Dynamic Programming, which consist of an iterative process over all MDP's states until convergence. The MDP for the traffic demand allocation problem consists of all the possible network topology states and the transition probabilities between states. Notice that in our scenario we have uniform transition probabilities from one state to the next. One limitation of solving MDPs \emph{optimally}  is that it becomes infeasible for large and complex optimization problems. As the problem size grows, so does the MDP's state space, where the space complexity (in number of states) is $S \approx$ $O(N\textsuperscript{E})$, having $N$ as the number of different capacities a link can have and $E$ as the number of links. Therefore, to solve the MDP the algorithm will spend more time on iterating over all MDP's states.

\section{GNN-based DRL agent design}
\label{section:4}

In this section, we describe the DRL+GNN architecture proposed in this paper. On one side, we have the GNN-based DRL agent which defines the actions to apply on the network topology. These actions consist of allocating the demands on one of the candidate paths. Our DRL agent implements the DQN algorithm~\cite{mnih2013playing}, where the q-value function is modeled by a GNN. On the other side, we have an environment which defines the optimization problem to solve. This environment stores the network topology, together with the link features. In addition, the environment is responsible of generating the reward once an action is performed, which will indicate the agent if that action was good or not.

The learning process is based on an iterative process, where at each time step, the agent receives a graph-structured network state observation from the environment. Then, the GNN constructs a graph representation where the links of the topology are the graph entities. In this representation, the link hidden states are initialized considering the input link-level features and the routing action to evaluate (see more details in Sections~\ref{subsec:environment},~\ref{subsec:action-set} and~\ref{subsec:gnnarchi}). With this representation, an iterative message passing algorithm runs between the links' hidden states according to the graph structure. The output of this algorithm (i.e., new links hidden states) is aggregated into a global hidden state, that encodes topology information, and then is processed by a DNN. This process makes the GNN topology invariant because the global hidden state length is pre-defined and it will always have the same length for different topology sizes. At the end of the message passing phase, the GNN outputs a q-value estimate. This q-value is evaluated over a limited set of actions, and finally the DRL agent selects the action with highest q-value.

\subsection{Environment}
\label{subsec:environment}

The network state is defined by the topology links' features, which include the link capacity and link betweenness. The former indicates the amount of capacity available on the link. The latter is a measure of centrality inherited from graph theory that indicates how many paths may potentially traverse the link. From the experimental results we observed that this feature helps reduce the grid search of the hyperparameter tuning for the DRL agent. This is because the betweenness helps the agent converge faster to a good policy. In particular, we compute the link betweenness in the following way: for each pair of nodes in the topology, we compute $k$ candidate paths (e.g., the $k$ shortest paths), and we maintain a per-link counter that indicates how many paths pass through the link. Thus, the betweenness on each link is the number of end-to-end paths crossing the link divided by the total number of paths.

\subsection{Action Space}
\label{subsec:action-set}

In this section we describe how the routing actions are represented in the DRL+GNN agent. Note that the number of possible routing combinations for each source-destination node pair typically results in a high dimensional action space in large-scale real-world networks. This makes the routing problem  complex for the DRL agent, since it should estimate the q-values for all the possible actions to apply (i.e., routing configurations). To overcome this problem, the action space must be carefully designed to reduce the dimensionality. In addition, to enable generalization to other topologies, the action space should be equivalent across topologies. In other words, if the actions in the training topology are represented by shortest paths, in the evalutation topolgy they should also be shortest paths. If the action space would be different (e.g., multiple paths between a source-destination node pair), the agent will have problems learning and it will not generalize well. To leverage the generalization capability of GNN, we introduced the action into the agent in the form of a graph. This makes the action representation invariant to node and edge permutation, which means that, once the GNN is successfully trained, it is able to understand actions over arbitrary graph structures (i.e., over different network states and topologies).

Considering the above, we limit the action set to \textit{k} candidate paths for each source-destination node pair. To maintain a good trade-off between flexibility to route traffic and the cost to evaluate all the possible actions, we selected a set with the \textit{k=4} shortest paths (by number of hops) for each source-destination node pair. This follows a criteria originally proposed by~\cite{suarez2019routingJ}. Note that the action set differs depending on the source and destination nodes of the traffic demand to be routed.

To represent the action, we introduce it within the network state. Particularly, we consider an additional link-level feature, which is the bandwidth allocated over the link after applying the routing action. This value corresponds to the bandwidth demand of the current traffic request to be allocated. Likewise, the links that are not included in the path selected by the action will have this feature set to zero. Since our OTN environment has a limited number of traffic requests with various discrete bandwidth demands, we represent the bandwidth allocated with a \textit{N}-element one-hot encoding vector, where \textit{N} is the vector length.

\begin{figure}[!t]
  \centering
  \includegraphics[width=0.7\linewidth]{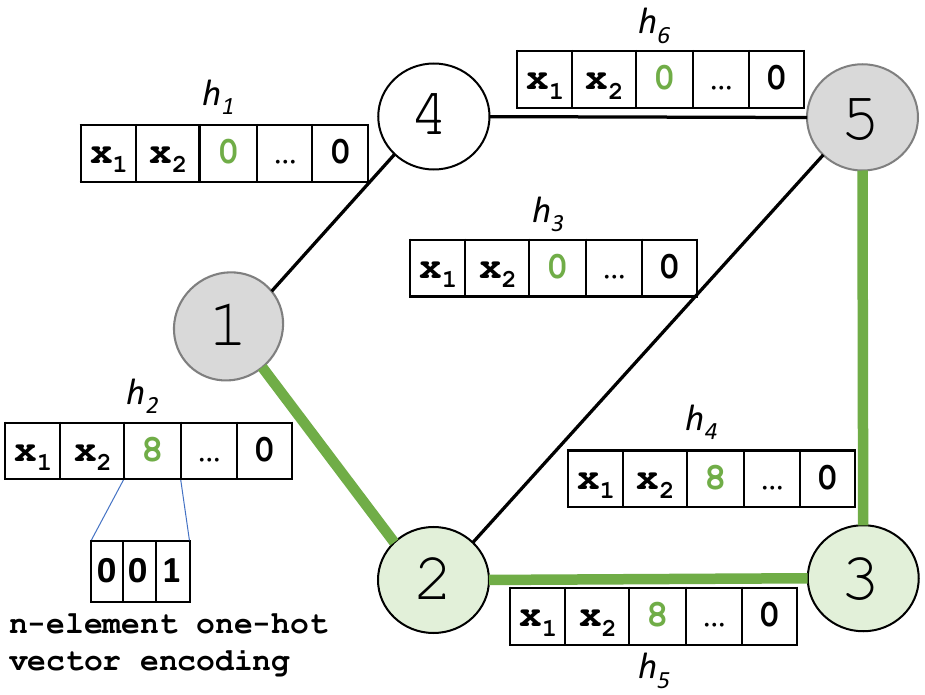}
  \caption{Action representation in the link hidden states.}
  \label{fig:actionSet}
\end{figure}

Figure~\ref{fig:actionSet} illustrates the representation of the action in the hidden state of the links in a simple network scenario. A traffic request from node 1 to node 5, with a traffic demand of 8 bandwidth units, is allocated over the path formed by the nodes \textit{\{1,2,3,5\}}. To summarize, Table~\ref{table:features} provides a description of the features included in the links' hidden states. These values represent both the network state and the action, which is the input needed to model the q-value function $Q(s,a)$.

The size of the hidden states is typically larger than the number of features in the hidden states. This is to enable each link to store information of himself (i.e., his own initial features) plus the aggregated information coming from all the links’ neighbors (see Section~\ref{subsec:gnnarchi}). If the hidden state size is equal to the number of link features, the links won’t have space to store information about the neighboring links without losing information. This results in a poor graph embedding after the readout function. On the contrary, if the state size is very large, it can lead to a large GNN model, which can overfit to the data. A common approach is to set the state size larger than the number of features and to fill the vector with zeros.

\begin{table}[!t]
\centering
\begin{tabular}{ll}
\toprule
Notation & Description\\
\midrule
\midrule
 $x_{1}$ & Link available capacity  \\
 $x_{2}$ &  Link Betweenness\\
 $x_{3}$ & Action vector (bandwidth allocated) \\
 $x_{4}-x_{N}$ &  Zero padding \\
\bottomrule
\end{tabular}
\caption{Input features of the link hidden states. \textit{N} corresponds to the size of the hidden state vector.}\label{table:features}
\end{table}

\subsection{GNN Architecture}
\label{subsec:gnnarchi}

The GNN model is based on the Message Passing Neural Network \cite{gilmer2017neural} model. In our case, we consider the link entity and perform the message passing process between all links. We choose link entities, instead of node entities, because the link features are what define the OTN routing optimization problem. Node entities could be added when addressing an optimization problem that needs to incorporate node-level features (e.g., I/O buffer size, scheduling algorithm). Algorithm \ref{alg:mp} shows a formal description of the message passing process where the algorithm receives as input the links' features ($x_{l}$) and outputs a q-value (\textit{q}).

\begin{figure}[!b]
  \centering
  \includegraphics[width=0.86\linewidth]{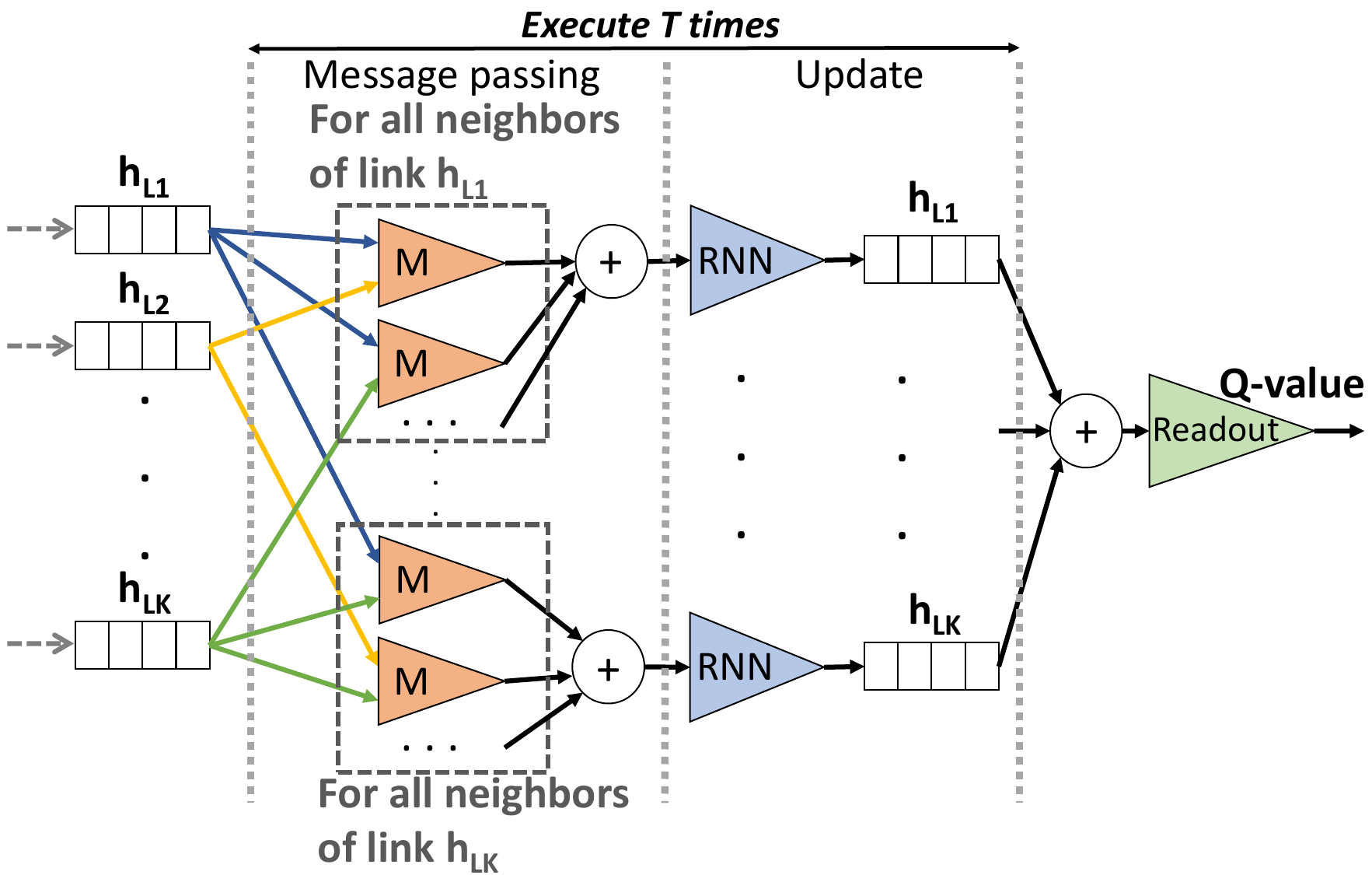}
  \caption{Message passing architecture.}
  \label{fig:mpnn}
\end{figure}

The algorithm performs \textit{T} message passing steps. A graphical representation can be seen in Figure~\ref{fig:mpnn}, where the algorithm iterates over all links of the network topology. For each link, its features are combined with those of the neighboring links using a fully-connected, corresponding to M in Figure~\ref{fig:mpnn}. The outputs of these operations are called \textit{messages} according to the GNN notation. Then, the \textit{messages} computed for each link with their neighbors are aggregated using an element-wise sum (line 5 in Algorithm~\ref{alg:mp}). Afterwards, a Recurrent NN (RNN) is used to update the link hidden states $h_{LK}$ with the new aggregated information (line 6 in Algorithm~\ref{alg:mp}). At the end of the message passing phase, the resulting link states are aggregated using an element-wise sum (line 7 in Algorithm~\ref{alg:mp}). The result is passed through a fully-connected DNN which models the readout function of the GNN. The output of this latter function is the estimated q-value of the input state and action.  

The role of the RNN is to learn how the link states change along the message passing phase. As the link information is being spread through the graph, each hidden state will store information from links that are farther and farther apart. Therefore, the concept of time appears. RNNs are a NN architecture that are tailored to capture sequential behavior (e.g., text, video, time-series). In addition, some RNN architectures (e.g., GRU) are designed to process large sequences (e.g., long text sentences in NLP). Specifically, they internally contain gates that are designed to mitigate the vanishing gradients, a common problem with large sequences~\cite{cho2014properties}. This makes RNNs suitable to learn how the links’ state evolve during the message passing phase, even for large \textit{T}. 

\subsection{DRL Agent Operation}
\label{subsec:drloperation}

The DRL agent operates by interacting with the environment. In Algorithm \ref{alg:operation} we can observe a pseudocode describing the DRL agent operation. At the beginning, we initialize the environment \textit{env} by initializing all the link features. At the same time, the environment generates a traffic demand to be allocated by the tuple $\{src, dst, bw\}$ and an environment state \textit{s}. We also initialize the cumulative reward to zero, define the action set size and create the experience replay buffer (\textit{agt.mem}). Afterwards, we execute a while loop (lines \ref{line:while-init}-\ref{line:while-end}) that finishes when there is some demand that cannot be allocated in the network topology. For each of the \textit{k=4} shortest paths, we allocate the demand along all the links forming the path and compute a q-value (lines \ref{line:get_path}-\ref{line:qvalue}). Once we have the q-value for each state-action pair, the next action \textit{a} to apply is selected using an \mbox{$\epsilon$-greedy} exploration strategy (line \ref{line:epsilon_greedy}) \cite{mnih2013playing}. The action is then applied to the environment, leading to a new state \textit{s'}, a reward \textit{r} and a flag \textit{Done} indicating if there is some link without enough capacity to support the demand. Additionally, the environment returns a new traffic demand tuple $\{src', dst', bw'\}$. The information about the state transition is stored in the experience replay buffer (line \ref{line:rememb}). This information will be used later on to train the GNN in the \textit{agt.replay()} call (line \ref{line:replay}), which is executed every $M$ training iterations.

\begin{algorithm}[!t]
\caption{Message Passing}
\begin{algorithmic}[1]
\Statex $\textbf{Input}: \mathbf x_l$
\Statex $\textbf{Output}: \mathbf h_l^T, q$
\ForEach{$ l \in \mathcal{L}$}\label{lin:loop1}
\State $h_l^0 \gets [\mathbf x_l,0\ldots, 0]$
\EndFor
\For{$t=1$ to $T$}
\ForEach{$ l \in \mathcal{L}$}\label{lin:h4}
\State $M_{l}^{t+1} = \sum_{i \in N(l)} m\left(h_{l}^{t}, h_{i}^{t}\right)$
\State $h_{l}^{t+1} = u\left(h_{l}^{t}, M_{l}^{t+1}\right)$
\EndFor
\EndFor
\State $rdt \gets \sum_{l \in \mathcal{L}} h_{l}$\label{lin:elemsum}
\State $q \gets R(\textit{rdt})$\label{lin:redout}
\end{algorithmic}
\label{alg:mp}
\end{algorithm}

\section{Experimental results}
\label{section:exper}

In this section we evaluate our GNN-based DRL agent to optimize the routing configuration in the OTN scenario described in Section~\ref{section:optimi}. Particularly, the experiments in this section are focused on evaluating the performance and generalization capabilities of the proposed DRL+GNN agent. 
Afterwards, in Section~\ref{sec:discussion}, we analyze the scalability properties of our solution and discuss other relevant aspects related to the deployment on production networks.

\begin{algorithm}[!b]
\caption{DRL Agent operation}
\begin{algorithmic}[1]
\State $s, src, dst, bw \gets \textit{env.init\_env()}$
\State $reward \gets 0$, $k \gets 4$, $agt.mem \gets \textit{\{ \}}$, $Done \gets \textit{False}$
\While {not Done}\label{line:while-init}
\State $k\_q\_values \gets \textit{\{ \}}$
\State $k\_shortest\_paths \gets \textit{compute\_k\_paths(k, src, dst)}$\label{line:kpaths}
\For {\textit{i} in \textit{${0,...,k}$}}
\State $p' \gets \textit{get\_path(i, k\_shortest\_paths)}$\label{line:get_path}
\State $s' \gets \textit{env.alloc\_demand(s, p', src, dst, dem)}$
\State $k\_q\_values[i] \gets \textit{compute\_q\_value(s', p')}$ \label{line:qvalue}
\EndFor
\State $q\_value \gets \textit{epsilon\_greedy(k\_q\_values, $\epsilon$)}$\label{line:epsilon_greedy}
\State $a \gets \textit{get\_action(q\_value, k\_shortest\_paths, s)}$
\State $r,Done,s',src',dst',bw' \gets \textit{env.step(s, a)}$
\State $agt.rmb(s, src, dst, bw, a, r, s', src', dst', bw')$\label{line:rememb}
\State $reward \gets reward + r$
\State {$\textbf{If} \textit{~training\_steps \% M} == \textit{0:}$ \textbf{agt.replay()}}\label{line:replay}
\State $src \gets \textit{src'}; dst \gets \textit{dst'}; bw \gets \textit{bw'}, s \gets \textit{s'}$\label{line:while-end}
\EndWhile 
\end{algorithmic}
\label{alg:operation}
\end{algorithm}

\subsection{Evaluation Setup}

We implemented the DRL+GNN solution described in Section~\ref{section:4} with Tensorflow~\cite{abadi2016tensorflow} and
evaluated it on an OTN network simulator implemented using the OpenAI Gym framework~\cite{1606.01540}. The source code, together with all the training and evaluation results are publicly available~\cite{code}.

In the OTN simulator, we consider three traffic demand types (ODU2, ODU3, and ODU4), whose bandwidth requirements are expressed in terms of multiples of ODU0 signals (i.e., 8, 32, and 64 ODU0 bandwidth units respectively)~\cite{itu}. When the DRL agent correctly allocates a demand, it receives an immediate reward being the bandwidth of the current traffic demand if it was properly allocated, otherwise the reward is 0. We consider that a demand is successfully allocated if all the links in the path selected by the DRL agent have enough available capacity to carry such demand. Likewise, episodes end when a traffic demand was not correctly allocated. Traffic demands are generated by uniformly selecting a source-destination node pair and a traffic demand type (ODUk).
This makes the problem even more difficult for the DRL agent, since the uniform traffic distribution hinders the exploitation of prediction systems to anticipate possible demands difficult to allocate. In other words, all traffic demands are equally probable to appear in the future, making it more difficult for the DRL agent to estimate the expected future rewards.

Initial experiments were carried out to choose an appropriate gradient-based optimization algorithm and to find the hyperparameter values for the DRL+GNN agent. For the GNN model, we defined the links' hidden states $h_{l}$ as 27-element vectors (filled with the features described in Table~\ref{table:features}). Note that the size of the hidden states is related to the amount of information they may potentially encode. Larger network topologies and complex network optimization scenarios might need larger sizes for the hidden state vectors. In every forward propagation of the GNN we execute \textit{T=7} message passing steps using batches of 32 samples. The optimizer used is the Stochastic Gradient Descent~\cite{bottou2010large} with a learning rate of $10^{-4}$ and a momentum of 0.9. We start the \mbox{$\epsilon$-greedy} exploration strategy with $\epsilon$=1.0 and maintain this value during 70 training iterations. Afterwards, $\epsilon$ decays exponentially every episode. The experience buffer stores 4,000 samples and is implemented as a FIFO queue (first in, first out). We applied \textit{l2} regularization and dropout to the readout function with a coefficient of 0.1 in both cases. The discount factor $\gamma$ was set to 0.95.

\subsection{Methodology}

We divided the evaluation of our DRL+GNN agent in two sets of experiments. In the first set, we focused on reasoning about the performance and generalization capabilities of our solution. For illustration purposes, we chose two particular network scenarios and analyzed them extensively. As a baseline, we implemented the DRL-based system proposed in~\cite{suarez2019routingJ}, a state-of-the-art solution for routing optimization in OTNs. Later on, in Section~\ref{sec:discussion}, we evaluated our solution on real-world network topologies and analyzed its scalability in terms of computation time and generalization capabilities.

To find the \emph{optimal} MDP solution to the OTN optimization problem is infeasible due to its complexity. Take as an example  a  small  network topology with 6 nodes and 8 edges, where the links have capacities of 3 ODU0 units, there is only one bandwidth type available (1 ODU0) and there are 4 possible actions. The resulting number of states of the MDP is 5\textsuperscript{$8$}*$6$*$5$*$1 \approx 1.17e7$ . To find a solution to the MDP we can use Dynamic Programming algorithms such as value iteration. However, this algorithm has a time complexity to solve the MDP of $O(S\textsuperscript{2}A)$, where $S$ and $A$ are the number of states and actions respectively and $S \approx O(N\textsuperscript{E})$, having N as the number of different capacities a link can have and E as the number of links.

As an alternative, we compare the DRL+GNN agent performance with a theoretical fluid model (labeled as \textit{Theoretical Fluid}). This model is a theoretical approach which considers that traffic demands may be split into the \textit{k=4} candidate paths proportionally to the available capacity they have. This routing policy is aimed at avoiding congestion on links. For instance, paths with low available capacity will carry a small proportion of the traffic volume from new demands. Note that this model is non-realizable because ODU demands cannot be split in real OTN scenarios. However, this model is fast to compute and serves us as a reference to compare the performance of the DRL+GNN agent. In addition, we also use a load balancing routing policy (LB), which selects uniformly random one path among the \textit{k=4} candidate shortest paths to allocate the traffic demand.

We trained the DRL+GNN agent in an OTN routing scenario on the 14-node Nsfnet topology~\cite{hei2004wavelength}, where we considered that the links represent lightpaths with capacity for 200 ODU0 signals. Note that the capacity is shared on both directions of the links and that the bandwidth of different traffic demands is expressed in multiples of ODU0 signals (i.e., 8, 32 or 64 ODU0 bandwidth units). We ran 1,000 training iterations where the agent received traffic demands and allocated them on one of the \textit{k=4} shortest paths available in the action set. The model with highest performance was selected to be benchmarked against traditional routing optimization strategies and state-of-the-art DRL-based solutions.

\subsection{Performance evaluation against state-of-the-art DRL-based solutions}

In this evaluation experiment, we compare our DRL+GNN agent against state-of-the-art DRL-based solutions. Particularly, we adapted the solution proposed in \cite{suarez2019routingJ} to operate in scenarios where links share their capacity in both directions. We trained two different instances of the state-of-the-art DRL agent in two network scenarios: 
the 14-node Nsfnet and the 24-node Geant2 topology~\cite{geant2}. 
We made 1,000 experiments with uniform traffic generation to provide representative results. Note that both, the proposed DRL+GNN agent and the state-of-the-art DRL solution, were evaluated over the same list of generated demands.

\begin{figure}[!t]
    \begin{subfigure}[]{0.495\columnwidth}
	\includegraphics[width=1.0\linewidth,height=3.4cm]{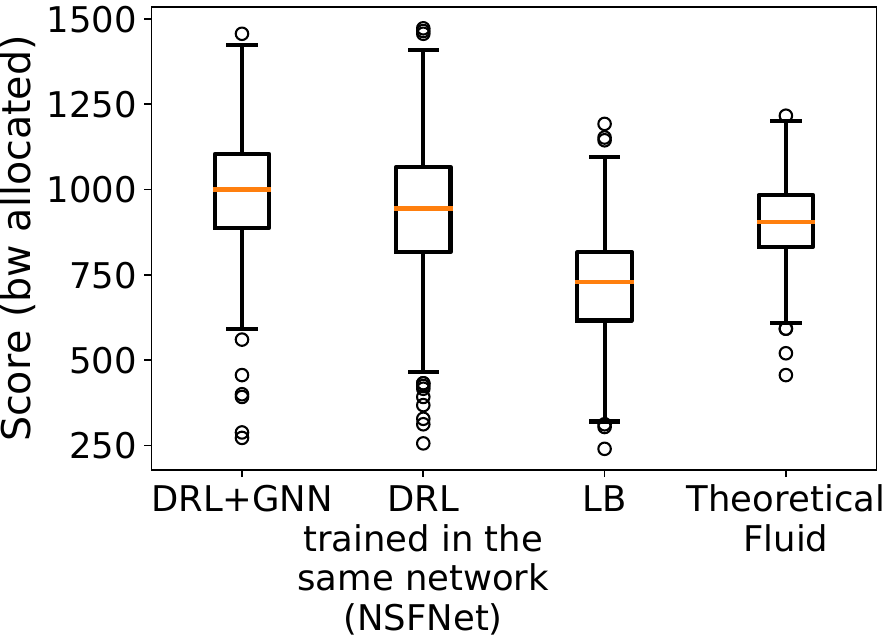}
        \caption{Evaluation on Nsfnet}
	\label{subfig:perfevalNsfnet}
    \end{subfigure}
    \begin{subfigure}[]{0.495\columnwidth}
	\includegraphics[width=1.0\linewidth,height=3.4cm]{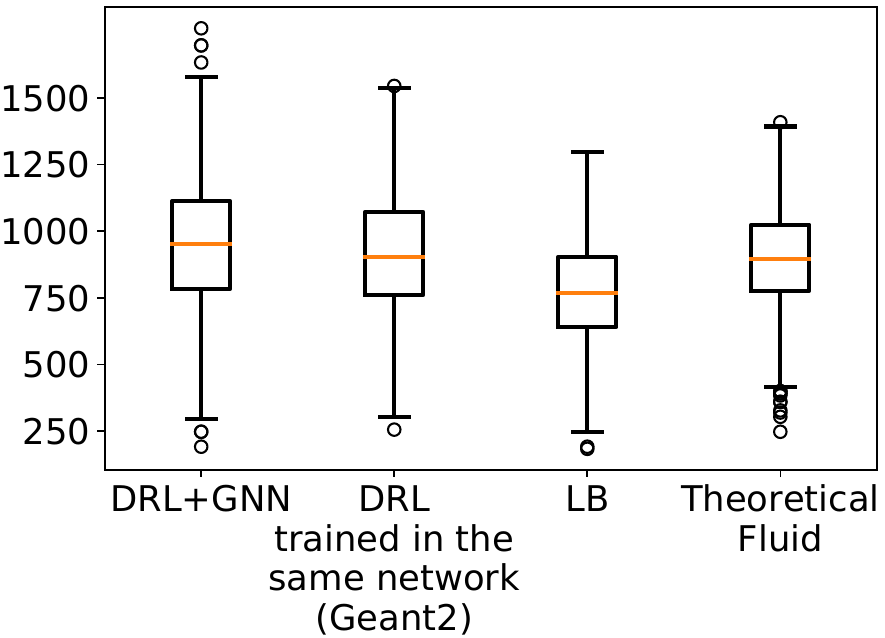}
        \caption{Evaluation on Geant2}
	\label{subfig:perfevalGeant2}
    \end{subfigure}
    
     \medskip
    
    \begin{subfigure}[]{0.495\columnwidth}
	\includegraphics[width=1.0\linewidth,height=3.4cm]{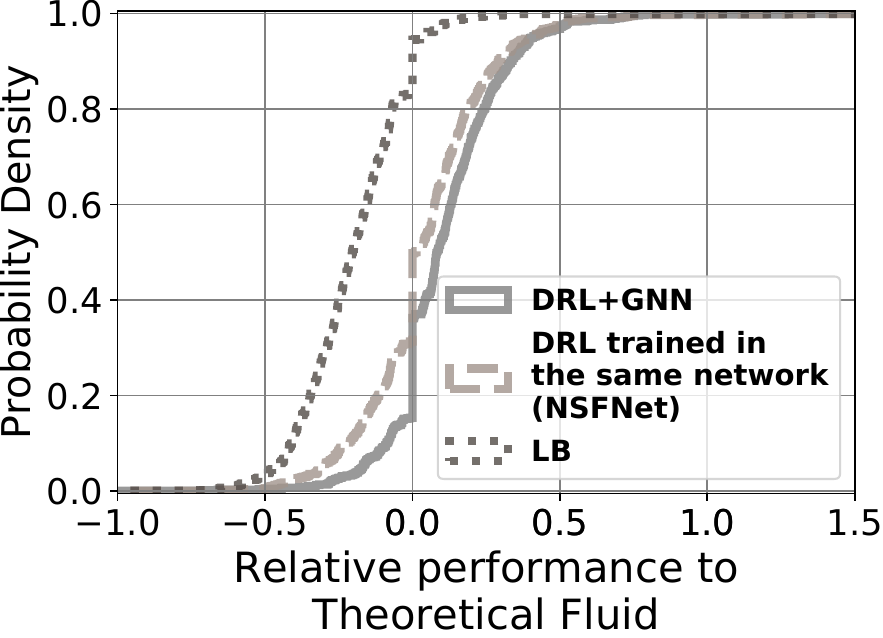}
        \caption{Evaluation on Nsfnet}
	\label{subfig:perfcdfNsfnet}
    \end{subfigure}
    \begin{subfigure}[]{0.495\columnwidth}
	\includegraphics[width=1.0\linewidth,height=3.4cm]{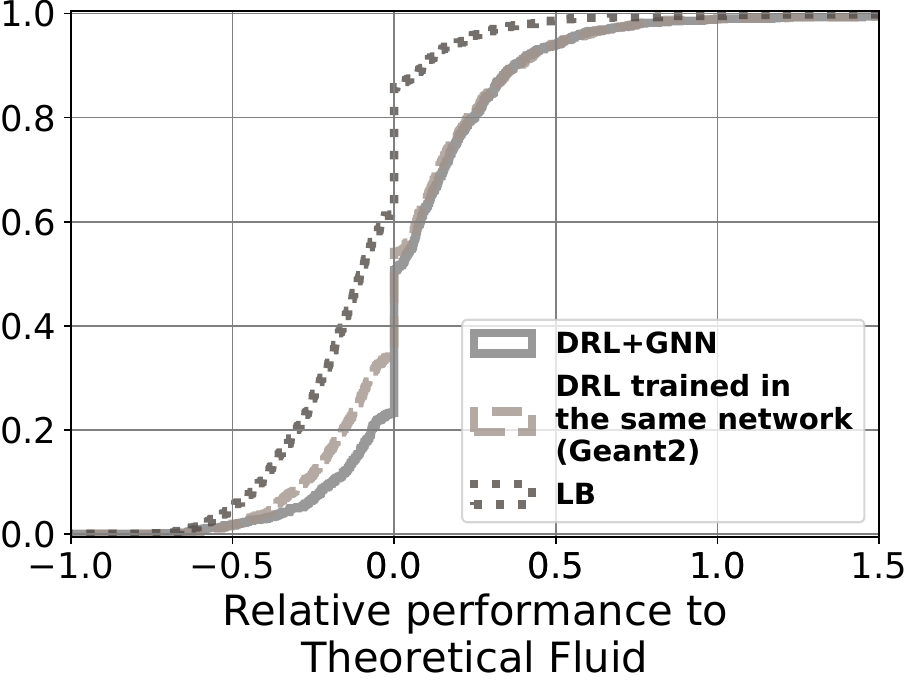}
        \caption{Evaluation on Geant2}
	\label{subfig:perfcdfGeant2}
    \end{subfigure}
     \caption{Performance evaluation against state-of-the-art DRL. Notice that the vertical lines in \ref{subfig:perfcdfNsfnet} and \ref{subfig:perfcdfGeant2} indicate the same performance as the theoretical fluid model.}\label{fig:boxplots}
\end{figure}

\begin{figure}[!t]
    \begin{subfigure}[]{0.495\columnwidth}
	\includegraphics[width=1.0\linewidth,height=3.4cm]{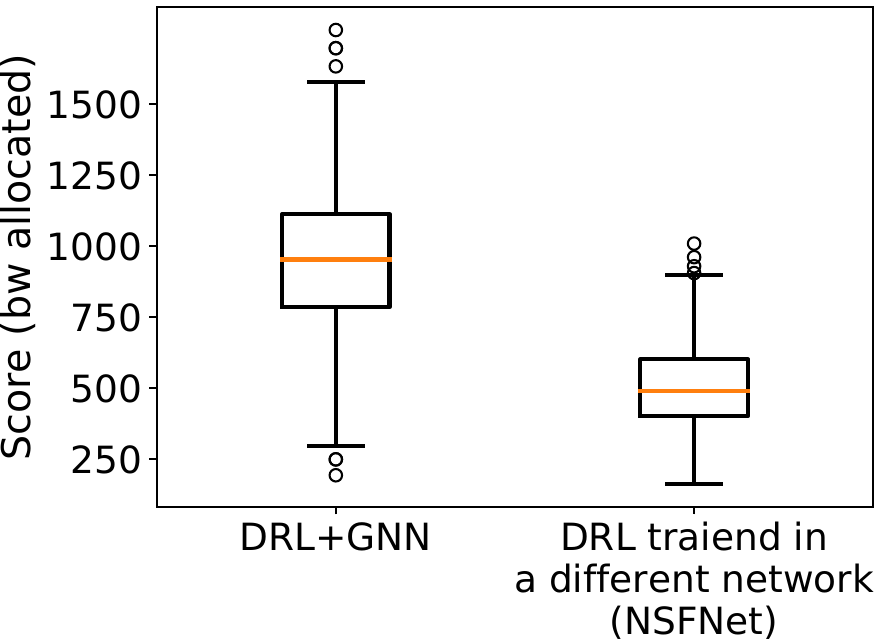}
    \caption{Bandwidth allocated}
	\label{subfig:genevalGeant2}
    \end{subfigure}
    \begin{subfigure}[]{0.49\columnwidth}
	\includegraphics[width=1.0\linewidth,height=3.4cm]{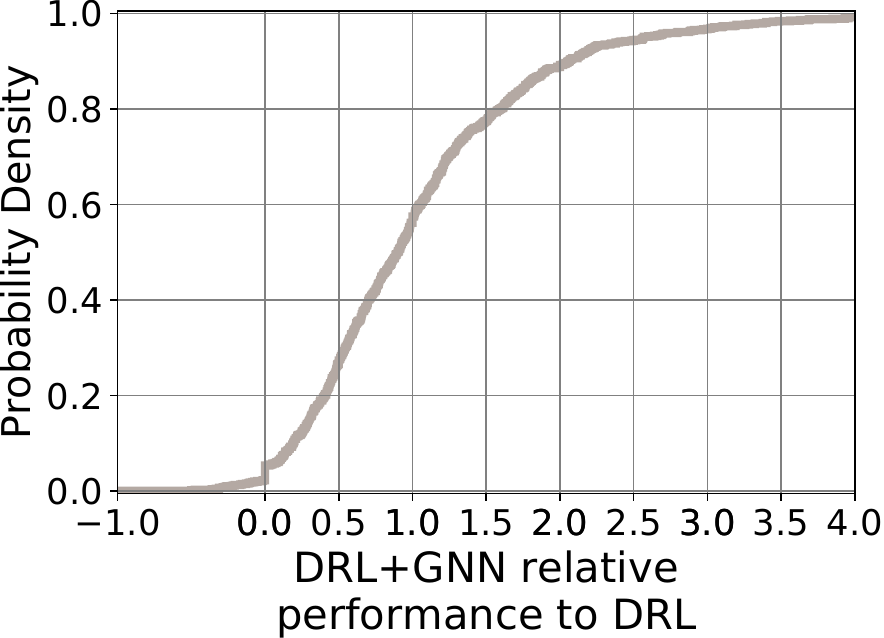}
        \caption{CDF}
	\label{subfig:gencdfevalGeant2}
    \end{subfigure}
     \caption{Evaluation on Geant2 of DRL-based solutions trained on Nsfnet.}\label{fig:genboxplots}
\end{figure}

We run two experiments to compare the \emph{performance} of our DRL+GNN with the results obtained by the state-of-the-art DRL (SoA DRL). In the first experiment, we evaluated the DRL+GNN agent against the SoA DRL agent trained on Nsfnet, the LB routing policy, and the theoretical fluid model. We evaluated the four routing strategies on the Nsfnet topology and compared their performance. In Figure~\ref{subfig:perfevalNsfnet}, we can observe a bloxplot with the evaluation results of 1,000 evaluation experiments. The y-axis indicates the agent score, which corresponds to the bandwidth allocated by the agent. Figure~\ref{subfig:perfcdfNsfnet} shows the Cumulative Distribution Function (CDF) of the relative score obtained with respect to the fluid model. In this experiment we could also observe that the proposed DRL+GNN agent slightly outperforms the SoA DRL-based by allocating 6.6\% more bandwidth. In the second experiment, we evaluated the same models (DRL+GNN, SoA DRL, LB, and Theoretical Fluid) on the Geant2 topology, but in this case the SoA DRL agent was trained on Geant2. The resulting boxplot can be seen in Figure~\ref{subfig:perfevalGeant2} and the CDF of the evaluation samples in Figure~\ref{subfig:perfcdfGeant2}. Similarly, in this case our agent performs slightly better than the SoA DRL approach (3\% more bandwidth).

We run another experiment to compare the \emph{generalization capabilities} of our DRL+GNN agent. In this experiment, we evaluated the DRL+GNN agent (trained on Nsfnet) against the SoA DRL agent trained on Nsfnet, and evaluated both agents on the Geant2 topology. The resulting boxplot can be seen in Figure~\ref{subfig:genevalGeant2} and the corresponding CDF in Figure~\ref{subfig:gencdfevalGeant2}. The results indicate that in this scenario the DRL+GNN agent also outperforms the SoA DRL agent. In this case, in 80\% of the experiments our DRL+GNN agent achieved more than 45\% of performance improvement with respect to the SoA DRL proposal. These results show that while the proposed DRL+GNN agent is able to generalize and achieve outstanding performance in the unseen Geant2 topology (Figure~\ref{subfig:genevalGeant2} and Figure~\ref{subfig:gencdfevalGeant2}), the SoA DRL agent performs poorly when applied to topologies not seen during training.  This reveals the lack of generalization capability of the latter DRL-based solution compared to the agent proposed in this paper.

\begin{figure}[!b]
    \begin{subfigure}[]{0.495\columnwidth}
	\includegraphics[width=1.0\linewidth,height=3.4cm]{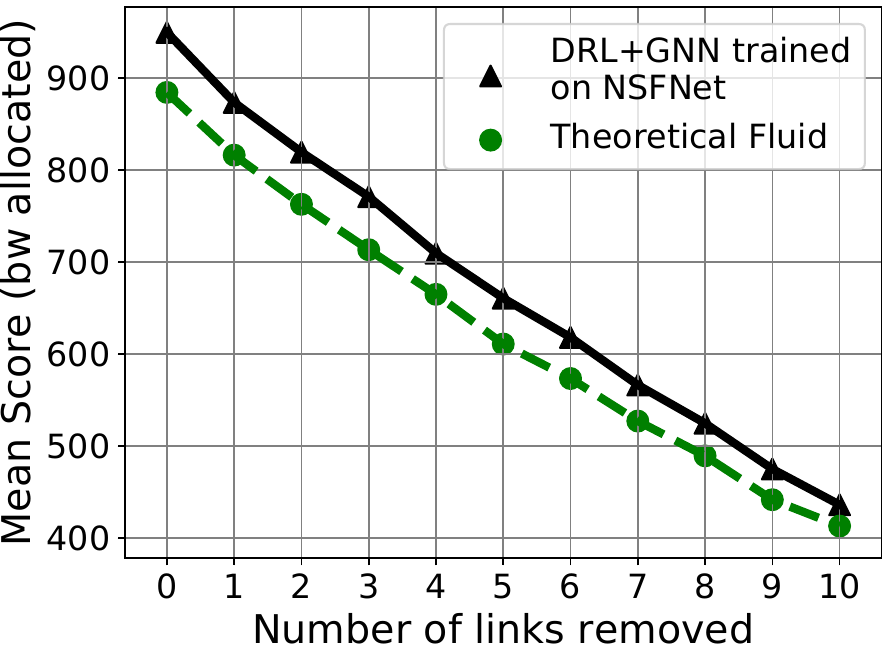}
        \caption{}
	\label{subfig:linkfailSc}
    \end{subfigure}
    \begin{subfigure}[]{0.495\columnwidth}
	\includegraphics[width=1.0\linewidth,height=3.4cm]{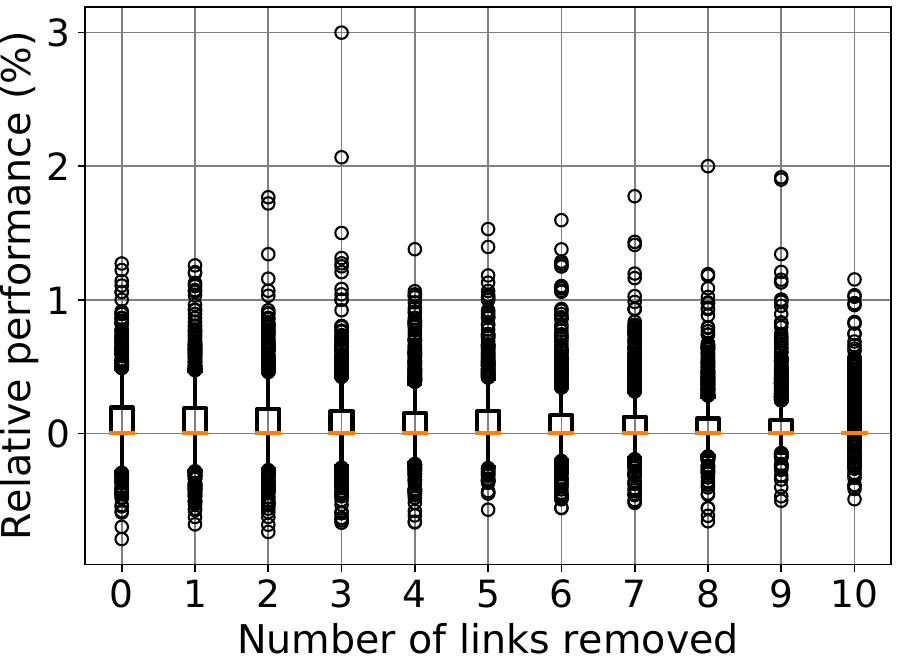}
        \caption{}
	\label{subfig:linkfailBox}
    \end{subfigure}
     \caption{DRL+GNN evaluation on a use case with link failures.}
\end{figure}

\subsection{Use case: Link failure resilience}
\label{subsec:linkfailure}

This subsection presents a use case where we evaluate if our DRL+GNN agent can adapt successfully to changes in the network topology. For this, we consider the case of a network with link failures. Previous work showed that real-world network topologies change during time (e.g., due to link failures) \cite{francois2005achieving, hartert2015declarative, jain2013b4}. These changes in network connectivity are unpredictable and they have a significant impact in protocol convergence \cite{francois2005achieving} or on fulfilling network optimization goals\cite{hartert2015declarative}. 

In this evaluation, we considered a range of scenarios that can experience up to 10 link failures. Thus, the DRL+GNN agent is tasked to find new routing configurations that avoid the affected links while still maximizing the total bandwidth allocated. We executed experiments where \mbox{$n\in[1,10]$} links are randomly removed from the Geant2 topology. We compare the score (i.e., bandwidth allocated) achieved by the DRL+GNN agent with respect to the theoretical fluid model. Figure~\ref{subfig:linkfailSc} shows the average score over 1,000 experiments (y-axis) as a function of the number of link failures (x-axis). There, we can observe that the DRL+GNN agent can maintain better performance than the theoretical baseline even in the extreme case of 10 concurrent link failures. Likewise, Figure~\ref{subfig:linkfailBox} shows the relative score of our DRL+GNN agent against the theoretical fluid model. In line with the previous results, the relative score is maintained as links are removed from the topology. This suggests that the proposed DRL+GNN architecture is able to adapt to topology changes.

\begin{figure}[!t]
  \centering
  \includegraphics[width=0.8\linewidth]{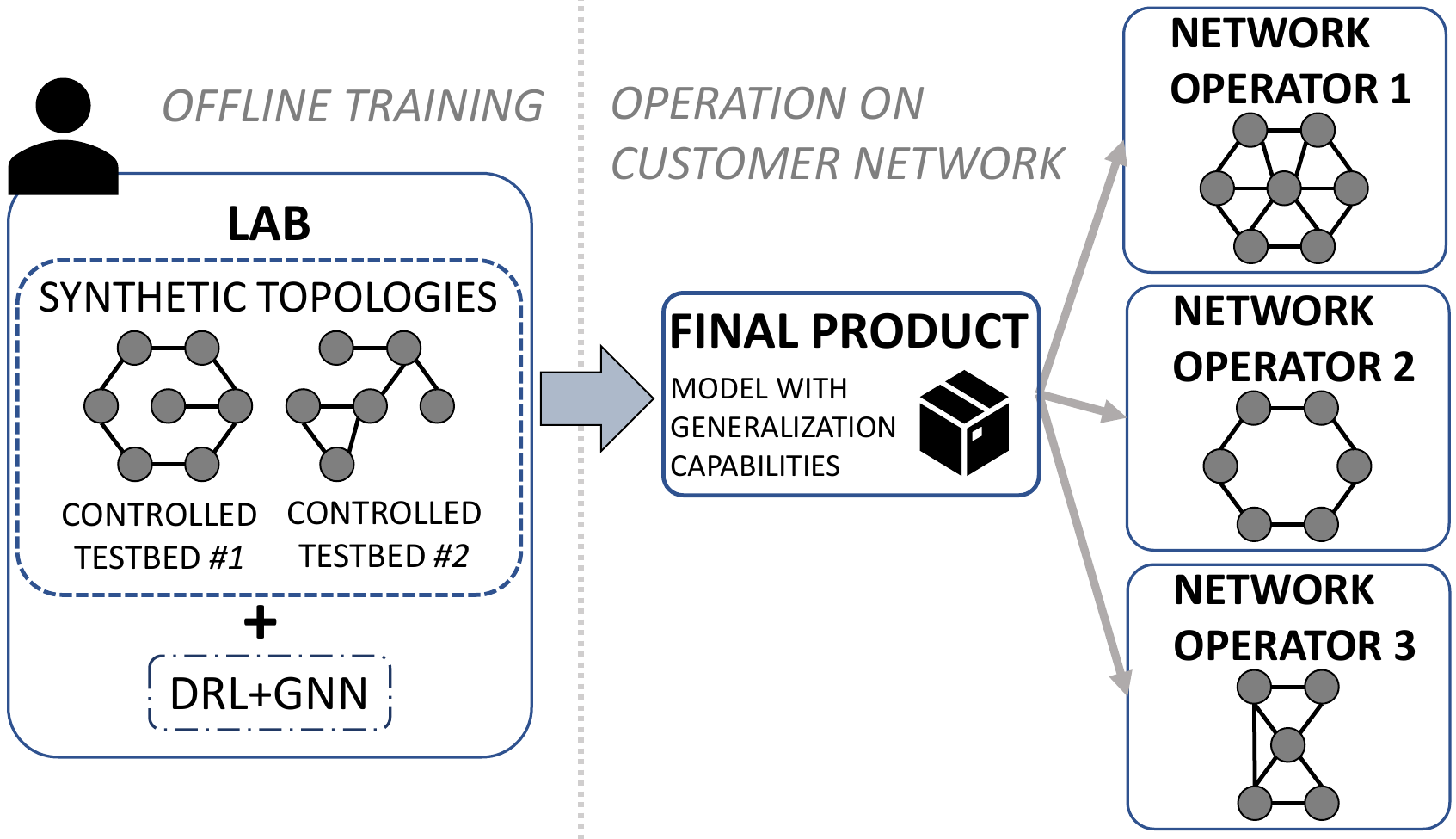}
  \caption{DRL+GNN deployment process overview by incorporating it into a product.}
  \label{fig:finalProduct}
\end{figure}

\section{Discussion on deployment}
\label{sec:discussion}

In this section we analyze and discuss relevant aspects of our DRL+GNN architecture towards deployment in production networks. In the context of self-driving networks, DRL cannot succeed without generalization capabilities. Training a DRL agent requires instrumenting the network with configurations that may disrupt the service. As a result, training in the customer's network may be unfeasible. With generalization capabilities, the DRL agent can be trained in a controlled lab (for instance at the vendor's facilities) and shipped to the customer. Once deployed, it can operate efficiently in an unseen network or scenario. Figure~\ref{fig:finalProduct} illustrates this training and deployment process of a product based on our DRL+GNN architecture.

To better understand the technical feasibility and scalability properties of such a product in terms of cost and generalization, we designed two experiments.
First, we analyze how the effectiveness of our agent scales with the network size, by training it in a single (small) network and evaluating its performance in synthetic and real-world network topologies. Second, we analyze the scalability of our agent in terms of computation time after deployment (i.e., the time it takes for the agent to make routing decisions). This is particularly relevant in real-time networking scenarios.

\begin{figure}[!t]
  \centering
  \begin{subfigure}[]{0.49\columnwidth}
  \includegraphics[width=1.0\linewidth]{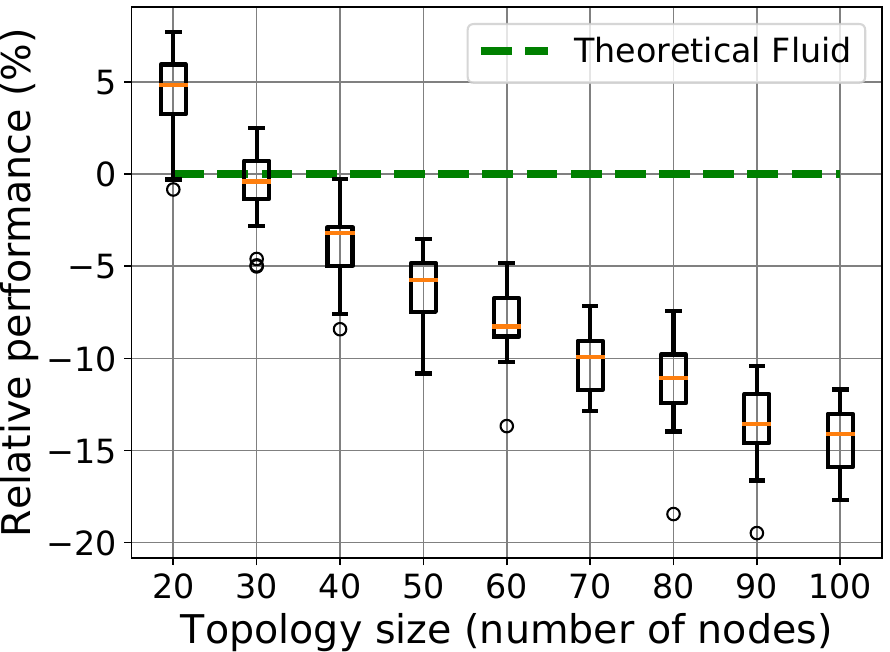}
  \caption{}
    \label{fig:synthTopoEvalBox}
   \end{subfigure}
    \begin{subfigure}[]{0.49\columnwidth}
  \includegraphics[width=1.0\linewidth]{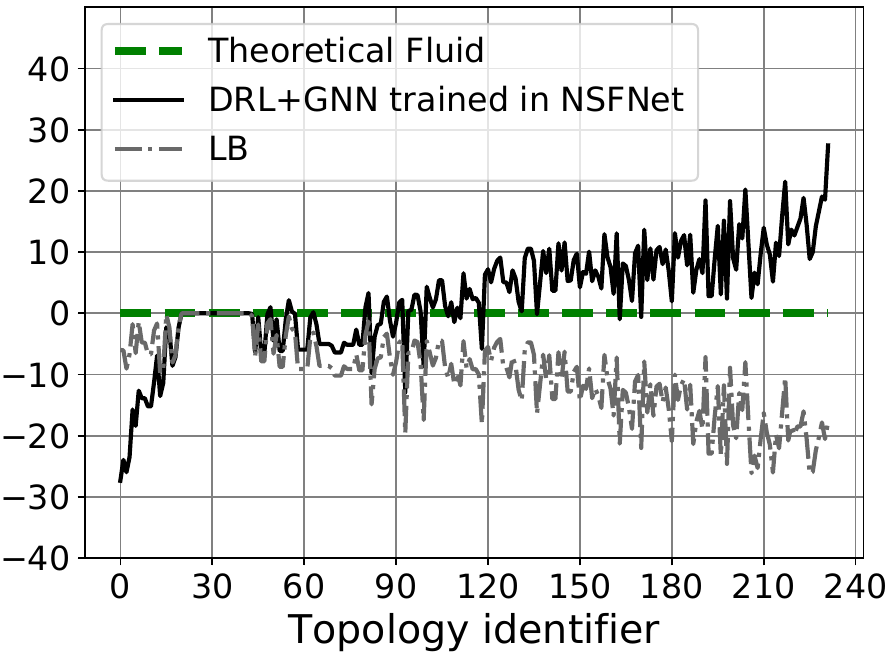}
    \caption{}
    \label{fig:generalization}
   \end{subfigure}
   \caption{DRL+GNN relative performance with respect to the fluid model over 180 synthetic topologies (a) and 232 real-world topologies (b).}
   \label{fig:figure8}
\end{figure}

\subsection{Generalization over network topologies}
\label{subsec:generalizationtopologies}

In both scenarios we used the DRL+GNN agent trained in a \emph{single} topology (14 nodes Nsfnet) and we analyzed its performance in larger topologies (up to 100 nodes) not seen during training.

\subsubsection{Synthetic topologies}
\label{subsubsection:syntht}

In this experiment we generated a total of 180 synthetic topologies with an increasing number of nodes. For each topology size --in number of nodes-- we generated 20 topologies and we evaluated the agent on 1,000 episodes. To do this, we used the NetworkX python library \cite{hagberg2008exploring} to generate random network topologies between 20 and 100 nodes with similar \textit{average node degree} to Nsfnet. This allows us to analyze how the network size affects the performance.

Figure~\ref{fig:synthTopoEvalBox} shows how the performance scales inversely with the topology size. For benchmark purposes, we computed the relative score with respect to the theoretical fluid model. The agent shows a remarkable performance in unseen topologies. As an example, the agent has a similar performance to the theoretical fluid model in the 30-node topologies, which double the size of the single 14-node topology seen during training. In addition, in the 100-node topologies, we observe only a 15\% drop in performance. This result shows that the generalization properties of our solution degrade gracefully with the size of the network. It is well-known that deep learning models lose generalization capability as the distribution of the data seen during training differs from the evaluation samples (see Section\ref{subsec:discussion}).

\subsubsection{Real-world network topologies}
\label{subsubsection:realworld}

\begin{table}[!b]
\centering
\begin{tabular}{lll}
\toprule
Feature & Minimum & Maximum\\
\midrule
\midrule
 Num. Nodes & 6 & 92 \\
 Num. Edges & 5 & 101\\
 Avg. node degree & 1.667 & 8\\
 Var. node degree &  0.001 & 41.415 \\
 Diameter &  1 & 31\\
\bottomrule
\end{tabular}
\caption{Real-world topology features (minimum and maximum values).}\label{table:topologiesZoo}
\end{table}

In this section we evaluate the generalization capabilities of our DRL+GNN agent, trained in Nsfnet, on 232 real-world topologies obtained from the Topology Zoo \cite{knight2011internet} dataset. Specifically, we take all the topologies that have up to 100 nodes. In Table \ref{table:topologiesZoo} we can see the features extracted from the resulting topologies. The diameter feature corresponds to the maximum eccentricity (i.e., maximum distance from one node to another node). The ranges of the different topology features indicate that our topology dataset contains different topology distributions. 

We executed 1,000 evaluation episodes and computed the average reward achieved by the DRL+GNN agents, the LB, and the theoretical fluid routing strategies for each topology. Then, we computed the relative performance (in \%) of our agent and the LB policy with respect to the theoretical fluid model. Figure~\ref{fig:generalization} shows the results where, for readability, we sort the topologies according to the difference of score between the DRL+GNN agent and the LB policy. In the left side of the figure we observe some topology samples where the scores of all three routing strategies coincide. This kind of behavior is normal in topologies where for each input traffic demand, there are not many paths to route the traffic demand (e.g., in ring or star topologies). As the number of paths increases, routing optimization becomes necessary to maximize the number of traffic demands allocated.

We also trained a DRL+GNN agent only in the Geant2 topology. The mean relative score (with respect to the theoretical fluid) of evaluating the model on all real-world topologies was +4.78\%. In the interest of space, we omit this figure. These results indicate that our DRL+GNN architecture generalizes well to topologies never seen during training, independently of the topology used during training.

These experiments show the robustness of our architecture to operate in real-world topologies that largely differ from the scenarios seen during training. Even when trained in a single 14-node topology, the agent achieves good performance in topologies of up to 100 nodes.

\subsection{Computation Time}
\label{subsec:computime}

In this section we analyze the computation time of an already trained DRL+GNN agent when deployed in a realistic scenario. For this purpose, we used the synthetic topologies generated before in Section~\ref{subsubsection:syntht}, and we executed 1,000 episodes for each one and we measured the computation time. This is the time the agent takes to select the best path to allocate all the incoming traffic requests. For this experiment we used \textit{off-the-shelf} hardware without any specific hardware accelerator (64-bit Ubuntu 16.04 LTS with processor Intel Core i5-8400 with 2.80GHz × 6 cores and 8GB of RAM memory).
Results should be understood only as a reference to analyze the scalability properties of our solution. Real implementations in a network device would be highly optimized.

\begin{figure}[!t]
  \centering
  \includegraphics[width=0.55\linewidth,height=3.2cm]{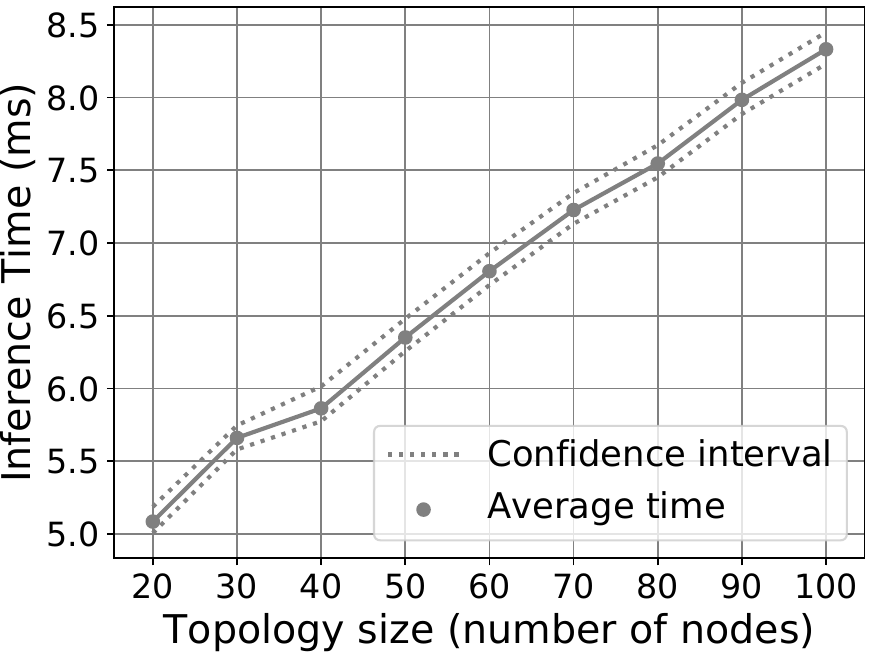}
  \caption{DRL+GNN average computation time (in milliseconds) over different topology sizes.}
  \label{fig:acttime}
\end{figure}

Figure~\ref{fig:acttime} shows the computation time for all episodes. The dots correspond to the average agent operation time over all the episodes and the confidence interval corresponds to the 5/95 percentiles. The execution time is in the order of few \textit{ms} and grows linearly with the size of the topology. This is expected due to the way the message-passing in the GNN has been designed. The results indicate that, in terms of deployment, the proposed DRL+GNN agent has interesting features. It is capable of optimizing unseen networks achieving good performance, as optimization algorithms, but in one single step and in tens of milliseconds, as heuristics.

\subsection{Discussion}
\label{subsec:discussion}

\begin{table}[!b]
\resizebox{\linewidth}{!}{
    \centering
    \begin{tabular}{|m{1.2cm}|m{0.9cm}|m{0.9cm}|m{0.95cm}|m{0.95cm}|m{1.2cm}|}
    \hline
        Topology Size & Mean Node Degree & Var. Node Degree & Node Betwee. & Edge Betwee. & DRL+GNN Perf. w.r.t. Fluid (\%) \\ \hline
        \textbf{Nsfnet (training)} & \textbf{3} & \textbf{0.2857} & \textbf{0.0952} & \textbf{0.1020} & \textbf{-} \\ \hline
        20 Nodes & 2.90 & 0.1050 & 0.1036 & 0.0988 & 4.305 \\ \hline
        30 Nodes & 2.93 & 0.0956 & 0.0844 & 0.0764 & -0.649 \\ \hline
        40 Nodes & 2.95 & 0.1025 & 0.0704 & 0.0623 & -3.945 \\ \hline
        50 Nodes & 2.96 & 0.1104 & 0.0620 & 0.0538 & -6.422 \\ \hline
        60 Nodes & 2.97 & 0.1056 & 0.0559 & 0.0476 & -8.103 \\ \hline
        70 Nodes & 2.97 & 0.0920 & 0.0522 & 0.0437 & -10.064 \\ \hline
        80 Nodes & 2.98 & 0.0956 & 0.0474 & 0.0395 & -11.380 \\ \hline
        90 Nodes & 2.98 & 0.1062 & 0.0436 & 0.0361 & -13.610 \\ \hline
    \end{tabular}
    }
\caption{Features for the Synthetic network topologies. The values correspond to the mean of all topologies from each topology size. As a reference, the first row corresponds to the Nsfnet topology used during training.}\label{table:synth_feat}
\end{table}

In this paper we propose a data-driven solution to solve a routing problem in OTN. This means that our DRL agent learns from data that is obtained from past interactions with the environment. This method has the main limitation that when evaluated on out-of-distribution data, its performance is expected to drop. In our scenario, out-of-distribution is any data related to network topology, link features and traffic matrix that is radically different from the data seen during the training process.

The experimental results on synthetic and real-world topologies (Section~\ref{subsubsection:syntht} and Section~\ref{subsubsection:realworld} respectively) show that the DRL+GNN architecture has performance issues on some topologies. This performance drop is related to the diverging network characteristics from the topology used during training. The link features are normalized and the traffic demands always have the same bandwidth values, which excludes them as the source of the performance drop. However, the network topology changes, which has a direct impact on the DRL agent performance.

\begin{table}[!t]
\resizebox{\linewidth}{!}{
    \centering
    \begin{tabular}{|m{1.2cm}|m{0.9cm}|m{0.9cm}|m{0.95cm}|m{0.95cm}|m{1.2cm}|}
    \hline
        Topology Id & Avg Node Degree & Var Node Degree & Node Betwee. & Edge Betwee. & DRL+GNN Perf. w.r.t. Fluid (\%) \\ \hline
        \textbf{Nsfnet (training)} & \textbf{3} & \textbf{0.29} & \textbf{0.0952} & \textbf{0.1020} & -\\ \hline
        0 & 2.42 & 9.59 & 0.0410 & 0.0484 & -27.357 \\ \hline
        1 & 3.51 & 15.17 & 0.0447 & 0.0394 & -23.944 \\ \hline
        2 & 2.00 & 41.41 & 0.0180 & 0.0298 & -25.965 \\ \hline
        229 & 2.31 & 0.98 & 0.1294 & 0.1615 & 19.066 \\ \hline
        230 & 2.06 & 1.75 & 0.1140 & 0.1340 & 18.570 \\ \hline
        231 & 2.07 & 2.22 & 0.0994 & 0.1244 & 27.430 \\ \hline
    \end{tabular}
    }
\caption{Features for the real-world network topologies. The relative performance is the mean of 1,000 evaluation episodes. As a reference, the first row corresponds to the Nsfnet topology used during training.}\label{table:realworld}
\end{table}

Table~\ref{table:synth_feat} shows different topology metrics for each topology size (in number of nodes). The edge betweenness is computed in the following way: for each edge compute the sum of the fraction of all-pairs of shortest paths that pass through the edge, and then make the mean of all edges. This applies in a similar way to the node betweenness. In addition, the DRL+GNN's performance with respect to the Theoretical Fluid model is also shown. The values correspond to the means from evaluating on all network topologies for each topology size (i.e., the means from the results in Figure~\ref{fig:synthTopoEvalBox}).

Even though the synthetic topologies were generated in a way to have a similar node degree like Nsfnet, we can see that other metrics diverge as the topologies become larger. Specifically, the node and edge betweenness become smaller, which indicates that the pairs of shortest paths are more distributed. In other words, for small topologies the nodes and edges have proportionally more shortest paths crossing them than for larger ones. The network metrics clearly indicate that the more different the topologies are than Nsfnet, the worse is the DRL's performance.

Table~\ref{table:realworld} shows a similar table but for the real-world topologies. In this case, the performance results correspond to the means from the results in Figure~\ref{fig:generalization}. Following a similar reasoning, we can see that the real-world topologies where the DRL+GNN architecture achieves the worst performance are radically different from Nsfnet (i.e., top left topologies from Figure~\ref{fig:generalization}). In addition, we visualized the topologies with id 0, 1 and 2 and observed that they correspond to topologies that have some nodes with a very high connectivity (see the variance of the node degree in Table~\ref{table:realworld}). Similarly to the synthetic topologies, we again observe that the more different the topologies are than Nsfnet, the worse is the DRL+GNN's performance.

There are several things that could be done to improve the generalization capabilities for such topologies. A straightforward approach would be to incorporate topologies with different characteristics to the training set. In addition, the DRL+GNN architecture could be improved using fine-tuned traditional Deep Learning techniques (e.g., regularization, dropout). Finally, the work from \cite{xu2018powerful} suggests that aggregating the information of the neighboring links using a combination of mean, min, max, and sum of the links’ states improves generalization. We consider that improving the generalization is outside the scope of our work and we left it as future work.

\section{Related work}

Network optimization is a well-known and established topic whose fundamental goal is to operate networks efficiently. Most of the works in the literature use traditional methods to optimize a network state based on Integer Linear Programming (ILP) or Constraint Programming (CP). A relevant work is \cite{hartert2015declarative} where they convert high-level goals, indicated by the network operator, into valid network configurations using constraint programming. The authors from \cite{gong2013efficient} propose a solution based on ILP for multicast routing in OTN. In addition, they propose to use a heuristic based on genetic algorithms to improve the locally optimal solution and to reduce the computational complexity. Similar work using genetic algorithms are \cite{gong2012two, klinkowski2012elastic}. 

To find the optimal routing configuration from a given traffic matrix is a fundamental problem, which has been demonstrated to be NP-hard \cite{wang2011study, christodoulopoulos2011elastic}. In this context, several DRL-based solutions have been proposed to address routing optimization. In \cite{chen2018deep} they propose a DRL solution for spectrum assignment using Q-learning and convolutional NNs. Similarly, in \cite{suarez2019routingJ} they propose a more elaborated representation of the network state to help the DRL agent capture easily the singularities of the network topology. In~\cite{sun2021scaledrl} they propose a scalable method to solve Traffic Engineering problems with DRL in large networks. The work from~\cite{9109571} combines DRL with Linear Programming to minimize the utilization of the most congested link. In~\cite{troia2020deep} they propose and compare different DRL-based algorithms to solve a Traffic Engineering problem in SD-WAN.

However, most of the proposed DRL-based solutions fail to generalize to unseen scenarios. This is because they pre-process the data from the network state and present it in the form of fixed-size matrices (e.g., adjacency matrix of a network topology). Then, these representations are processed by traditional neural network architectures (e.g., fully connected, convolutional neural networks). These neural architectures are not suited to learn and generalize over data that is inherently structured as a graph. Consequently, state-of-the-art DRL agents perform poorly when they are evaluated in different topologies that were not seen during the training.

There have been several attempts to use GNN in the communication networks field. In~\cite{geyer2018learning} they use GNN to learn shortest-path routing and max-min routing in a supervised learning approach. In~\cite{zhu2021network} they combine GNN with DRL to solve a network planning problem. Another relevant work is the one from~\cite{bernardez2021machine} where they use a distributed setup of DRL agents to solve a Traffic Engineering problem in a decentralized way. The work from~\cite{rusek2020routenet} proposes to use GNN to predict network metrics and a traditional optimizer to find the routing that minimizes some network metrics (e.g., average delay). Finally, GNNs have been proposed to learn job scheduling policies in a data-center scenario without human intervention \cite{mao2019learning}.

\section{Conclusion}
\label{section:conc}

In this paper, we presented a DRL architecture based on GNNs that is able to generalize to unseen network topologies. The use of GNNs to model the network environment allows the DRL agent to operate in different networks than those used for training. We believe that the lack of generalization was the main obstacle preventing the use and deployment of DRL in production networks. The proposed architecture represents a first step towards the development of a new generation of DRL-based products for networking.

In order to show the generalization capabilities of our DRL+GNN solution, we selected a classic problem in the field of optical networks. This served as a baseline benchmark to validate the generalization performance of our architecture. Our results show that the proposed DRL+GNN agent is able to effectively operate in networks never seen during training. Previous DRL solutions based on traditional neural network architectures were not able to generalize to other topologies.

A fundamental challenge that remains to be addressed towards the deployment of DRL techniques for self-driving networks is their black-box nature. DRL does not provide guaranteed performance for all network scenarios and its operation cannot be understood easily by humans. As a result, DRL-based solutions are inherently complex to troubleshoot and debug by network operators. In contrast, computer networks have been built around well-understood analytical and heuristic techniques, and such mechanisms are based on well-known assumptions that perform reasonably well across different scenarios. 
Such issues are not unique to self-driving networks, but rather common to the application of machine learning to many critical use-cases, such as self-driving cars. 

\ifCLASSOPTIONcompsoc
  \section*{Acknowledgments}
\else
  \section*{Acknowledgment}
\fi

This publication is part of the Spanish I+D+i project TRAINER-A (ref.PID2020-118011GB-C21), funded by MCIN/ AEI/10.13039/501100011033. This work is also partially funded by the Catalan Institution for Research and Advanced Studies (ICREA) and the Secretariat for Universities and Research of the Ministry of Business and Knowledge of the Government of Catalonia and the European Social Fund. This work was also supported by the Polish Ministry of Science and Higher Education with the subvention funds of the Faculty of Computer Science, Electronics and Telecommunications of AGH University and by the PL-Grid Infrastructure.

\ifCLASSOPTIONcaptionsoff
  \newpage
\fi



%

\bibliographystyle{IEEEtran}
\bibliography{references}

%









\end{document}